\documentclass[11pt, a4paper]{article}

\newcommand{\be}{\begin{equation}}
\newcommand{\ee}{\end{equation}}

\usepackage{graphicx}
\usepackage{amsmath}
\usepackage{a4}
\usepackage[colorlinks=true,hyperindex=true]{hyperref}

\begin{document}

\begin{titlepage}
  {\vspace{-0.5cm} \normalsize}
 
  \begin{center}
    \begin{LARGE}
      \textbf{Magnetic imaging of shipwrecks.}\\

    \end{LARGE}
  \end{center}

  \baselineskip 20pt plus 2pt minus 2pt
  \begin{center}
    \textbf{      Chris~Michael}

  \end{center}
  
  \begin{center}
    \begin{footnotesize}
      \noindent 
       Theoretical Physics Division, Dept. of Mathematical Sciences,
      \\University of Liverpool, Liverpool L69 7ZL, UK\\
      \vspace{0.2cm}

    \end{footnotesize}
  \end{center}
  \vspace{2cm}

  \begin{abstract}
    \noindent\begin{tabular*}{1.\linewidth}{@{\extracolsep{\fill}}l}
      \hline
    \end{tabular*} 
    
 The ferromagnetic material in a shipwreck on the seabed causes a
modification to the earth's magnetic field which can be measured at the
surface.  Proton magnetometer measurements at the surface are used to
locate wrecks.  Here I discuss how to interpret such data to explore the
shape and orientation  of the shipwreck on the seabed. I give details of
how to  model shipwrecks and deduce the magnetic signal that results. I
also  discuss how to analyse data in a more general way.
 As examples, I present and analyse data on the shipwrecks of YSTROOM and
 BOUBOULINA (ex COLONEL LAMB).

    \noindent\begin{tabular*}{1.\linewidth}{@{\extracolsep{\fill}}l}
      \hline
    \end{tabular*} 
  \end{abstract}

\end{titlepage}


\section{Introduction}
 \label{chap:int}

 Ships are mostly made of steel (iron in earlier times). 
 Iron and steel (also nickel and cobalt) are ferromagnetic, which means
there is a strong induced  magnetism when placed in an external magnetic
field. The earth has a magnetic  field (this is what makes a magnetic
compass work) and so will cause such  an induced magnetism. The net
effect is to modify the earth's magnetic field  in the vicinity of the
iron. As an approximation, one can say that the magnetic  field of the
earth `prefers' to go through the iron - so the field strength  is
enhanced where the field enters and leaves the iron and is reduced at
the sides.  Since the earth's field (in Britain) is downward angled, the
 ferromagnetic material will tend to increase the field above (and a bit
to the  south) of the submerged material, while further away the surface
field can be  reduced.

 This modification to the earth's magnetic field is what allows 
detection at the surface of the presence of iron on the seabed. The
induced magnetism of  the ferromagnetic material at the seabed produces
an anomalous field which  decreases as the inverse cube of the distance.
However, a large object (a ship)  will have an anomalous field that can
be detected over a significant area of the  surface. This makes for an
efficient method to detect such shipwrecks. 

 Here I discuss whether it is possible and practical to deduce something
about the shape of the shipwreck from surface measurements of the
magnetic anomaly.  This will be especially useful for shipwrecks that
are mostly buried  below the seabed. To investigate this, I present a
discussion of the  modelling of a shipwreck and the consequences for the
magnetic anomaly.  This allows a discussion of  how best to interpret
readings from a magnetometer taken at the surface.

As an example of this approach, I present my data from several
shipwrecks in Liverpool Bay and discuss models that describe the surface
anomaly observed.  In this case I can also compare with observations,
from diving, of the seabed  wreck.

The mathematical details of interpreting the surface anomaly are
presented in  Appendices. Appendix~A evaluates the surface magnetic
field anomaly from a  collection of induced magnetic dipoles
representing the shipwreck. Appendix~B  discusses how such induced
dipoles can arise and what their orientation will be. Appendix~C gives a
discussion  of methods to analyse the surface distribution, both in
general and with some assumptions. Appendix~D reviews creating a data 
grid from scattered data.

\section{Surface magnetic measurements}
 \label{chap:surf}

 The earth's magnetic field has a strength and a direction. In the
northern  hemisphere it is is downward angled. This is called the angle
of `dip'  or inclination. It is about 67$^0$ to the horizontal in
Britain in 2010 
 \newline (see \url{http://www.geomag.bgs.ac.uk/images/fig3.pdf}).  
 The horizontal component points to `magnetic north'. The deviation from
 true north is called the `magnetic deviation'. In Britain in 2010 this 
deviation is small - just a few degrees. The field strength is around
49000 nT (nanoTesla) in Britain 
  \newline (see \url{http://www.geomag.bgs.ac.uk/images/fig4.pdf})

Magnetometers measure the strength of a magnetic field. Some types of
magnetometer  measure the component in a given direction. This can be
useful for a compass. In a  rocking boat, these components will change
with the boat's orientation. A more stable  quantity is the total
intensity. This can be determined by combining the output  from three
orthogonal directions (the sum of squares is needed) or by using a 
magnetometer which measures this directly - such as a proton
magnetometer. Here I will mainly consider data obtained with a proton 
magnetometer since this can be accurate (to within 1 nT) and  equipment
is available   designed for use at sea. Data can be obtained every few
seconds (typically 2 secs) with GPS position and magnetic intensity
logged. It may also  be feasible to log depth to the seabed (and
possibly also depth of the magnetometer head if this is allowed to sink)

To avoid detecting the magnetism of one's own vessel, a magnetometer head is 
towed behind. Since position is detected (by GPS) on the vessel itself, a 
correction needs to be made for the fact that the magnetometer signal is 
from a position that was the vessel's position a short time ago.

The magnetic anomaly from a seabed shipwreck depends on the depth and
the size  of the wreck. A coaster in 30 metres may have a signal as
small as a few  hundred nT whereas a several thousand ton ship can have
a signal of  several thousands of nT. The signal is significant (say 10
nT or more) over quite a large  area and this is why a proton
magnetometer is such a successful wreck finder. I will discuss in more 
detail the nature of the surface anomaly and whether it is possible to 
deduce more detail about the wreck - such as its orientation underwater.

\section{Induced magnetisation}
 \label{chap:induce}

Ferromagnetic materials will have an induced magnetism when placed in a
magnetic field.  For ferromagnetic materials such as soft
iron, this  induced magnetism is the main effect. As discussed later,
steel can become  semi-permanently magnetised (again by the earth's
magnetic field during  construction or from subsequent voyages). This 
permanent magnetism can also be represented in the same way as the induced
magnetism, by a dipole distribution.

In summary, the ferromagnetic material of the shipwreck  will, in the
earth's magnetic field, acquire  an induced magnetism. This can  be
represented by a distribution of magnetic dipoles. The dipoles have a 
strength and a direction. The task is to estimate this dipole
distribution  in the case of a shipwreck and then evaluate the
modification  of the earth's magnetic field at the  surface (called the 
surface anomaly) caused by  those induced dipoles. See Appendix~A for
details  of determining the surface anomaly given the dipole
distribution.

The simplest case, discussed in Appendix~B1, is of a spherically
symmetric  distribution of ferromagnetic material. This is a
mathematician's ship! Even  so, it can yield some insight since it can
be solved exactly. One finds that the  induced magnetic dipole is directed
along the earth's field and is actually  effectively located at the
centre of the sphere of material. The surface distribution  from this 
induced dipole is easy to calculate (see Appendix~A) and is illustrated
in  fig.~\ref{fig:sample}. This shows that the field intensity  is
increased in a region above the source  and is decreased (especially 
northwards) away from the source. If the angle of dip were $90^0$ (as 
at the north magnetic pole) the intensity would be increased inside a 
circle centred on the source and of radius $1.414 z$ where $z$ is the
depth  and  the intensity would be decreased outside that radius. In
Britain where the angle of dip  is quite large (around $67^0$) the
situation is similar but distorted  as shown by fig.~\ref{fig:sample}. 

 The most characteristic feature, for detection purposes, is that the
maximum  is slightly south of the source (4 metres in the example) and
the most negative value (the  minimum) is only 12\% of the magnitude of
the maximum and is situated further north (32 metres in the example) of
the source. 
  
\begin{figure}[t]
  \centering
  \includegraphics[width=.9\linewidth]{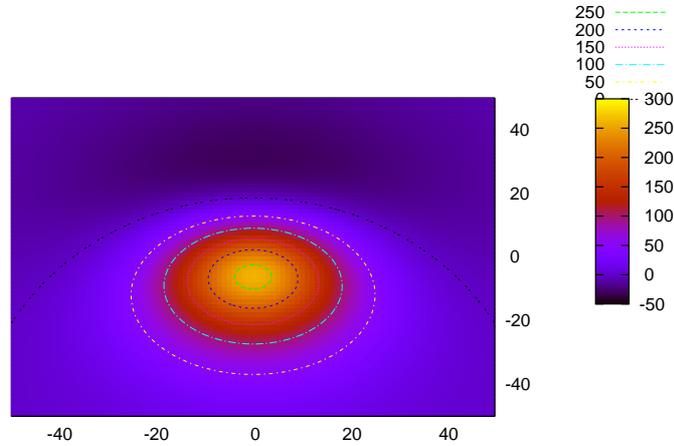}
  \caption{Magnetic field intensity anomaly (arbitrary units) at the
surface with distances in metres for a point magnetic source at the
origin and at depth 30 metres. Magnetic north is upwards, the angle  of
dip used is $67^0$.
  }
  \label{fig:sample}
\end{figure}

 If a shipwreck is modelled as a collection of spherical components
which  are not close to each other, then the surface distribution will
be given  by adding together the distributions from each of the
components. Although  this model would represent a scattered field of
cannon balls on the seabed, it is  actually a rather unrealistic way to
represent an iron shipwreck: rather there will  be flat steel plates
forming the hull, bulkheads and decks. 

I now discuss the induced magnetism in a rectangular steel plate (see
Appendix~B3). The essential feature is that the magnetic field has to
obey  certain conditions at the surface between the plate and the water.
These  imply that the induced dipoles must lie primarily oriented along
the plate  (i.e. tangential to the surface of the plate). This is a new
feature compared to the spherical situation:  the induced dipoles will
not necessarily be directed along the earth's magnetic field.

\begin{figure}[h]
  \centering
  \includegraphics[width=.9\linewidth]{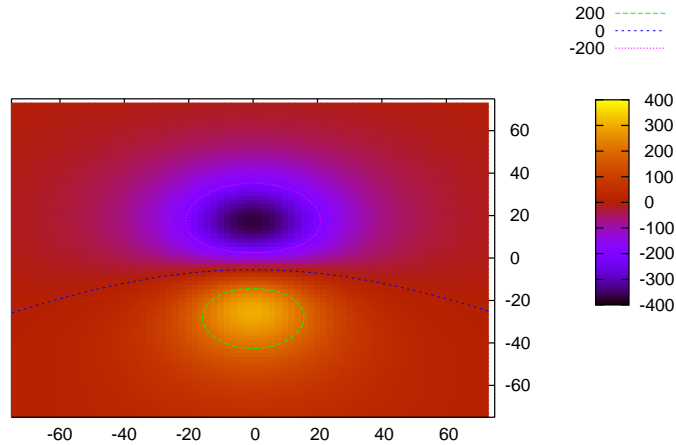}
  \caption{Magnetic field intensity anomaly (in nT) at the surface with
distances in metres for a horizontal flat rectangular plate (10m
$\times$ 40m, longest side lying N-S, thickness $t$ with $t
(\mu_r-1)=40$ metres) as magnetic source, centred  at the origin and at
depth 30 metres. Magnetic north is upwards, the angle  of dip used is
$67^0$. 
  }
  \label{fig:flat}  
\end{figure}

For example, a horizontal plate will thus have dipoles with no vertical
component and this  gives a very different surface distribution - see
fig.~\ref{fig:flat} for  an illustration. Here the negative peak
(anomaly reducing the intensity of the earth's field  at the surface) is
more prominent. Other cases are discussed in Appendix~B3.

Thus for a shipwreck made of iron/steel plates,  it is not likely, in
general,  that the induced magnetism will  lie along the earth's
magnetic field direction.

In general, the modification of the magnetic field intensity  at the
surface can be  evaluated given the distribution and orientation of
induced magnetic dipoles in the ferromagnetic  material of the shipwreck
(as described in Appendix~A). Approximately this will  be related to
the distribution of iron/steel underwater. 
 At large distances from the wreck, the net anomaly will have the form
of a  magnetic dipole centred on the wreck. The only information then
available will  be the total strength and the orientation of this
dipole. The total strength is a useful clue to the tonnage of the wreck.
 At first sight, the  dipole would be oriented along the earth's
magnetic field, so giving  little information about the wreck. This is
not actually  the case for a wreck containing steel (or iron) plates, so
some extra information  on the nature of the wreck can be gleaned.
 As I will show, if the depth is not much greater than the length of
the wreck, the  surface anomaly can give additional information about
the distribution of material  on the seabed.

 Given an extended source (big ship, etc) then one solves for the 
induced magnetism combining the contributions from all the iron/steel in
the wreck. This can be evaluated  by computer, in principle,  if one
knows that distribution: see Appendix~B4. However, if the  wreck has
collapsed, turned over while sinking, been partly demolished by
explosives, been salvaged, been moved by wave motion,...  one will not
have any accurate model of the distribution of ferromagnetic material.

In practice one will need to make a simplifying model of the wreck:  a
flat rectangular plate, a cylinder, a shoe box, a rectangular
cross-section  tube,....   

Some idea of the distribution of material on the seabed can be obtained
in  general, as described in Appendix~C, provided sufficiently precise
surface  data are available. This is discussed later. If one makes a
stronger assumption, that all the ferromagnetic material lies at the
same depth  and is all magnetised in the same direction, then it is
possible to extract the  material distribution on the seabed from the
surface measurement without constructing a model. This  is discussed in
Appendix~C1. One always can try this approach but if regions  on the
seabed come out with a negative amount of material, this shows that the
assumption was  not valid. This approach is related to the method known 
as `return to pole' which is described in Appendix~C2.

\section{Permanent magnetisation effects}
 \label{chap:permanent}

  What is quite well established is that a ship, a few years after
construction,  that voyages primarily in the northern hemisphere, will
have a semi-permanent  magnetisation in the vertical (downward)
direction. This semi-permanent magnetism arises from stresses in the
steel  of the vessel caused by wave motion, engine vibration, gunfire,
heat,  cargo loading, etc. It can change over a period of a year or so.
Ship's compasses are corrected for semi-permanent and induced magnetism
and this is a source of much expertise. Also detection of submarines  by
their magnetic signature is well studied.

 For example, submarines are regularly `depermed' - have their permanent
magnetisation  removed -  by using a strong oscillating magnetic field
that decreases  in strength. This reduces their magnetic `presence' so
they cannot be as easily  detected by magnetometers. During World War
II, a method of counteracting the magnetic `presence' of a ship  needed
to be developed - to avoid setting off magnetic mines. This involved a
permanent electric  current encircling the ship close to the waterline
which produced a vertical magnetic field which  cancelled out the 
semi-permanent magnetisation (and also the vertical component of the
induced magnetisation to some extent).  This was known as `degaussing'. 
Later on, deperming was also used.

A coda to this is that when  magnetic mines were made more
sensitive,  the degaussing currents were switched to maximise the ship's
magnetic signal -  causing the mines to explode some distance away from
the ship.

The magnetisation from construction (heating/cooling and hammering can
both give rise  to a permanent magnetisation) is not easy to predict.
What emerges from the reports on magnetic effects in real ships (and
submarines)  is that the vertical semi-permanent magnetic effect is
usually the dominant one after a few years at sea.
 What is quite unclear to me is what happens after a ship sinks. It no
longer  moves around and is not subject to vibration or wave slamming
(if sufficiently deep). So one possibility is that the pre-sinking
semi-permanent  magnetisation remains. Another possibility is that it
decays slowly and what remains is the induced magnetisation caused by
the earth's magnetic field direction  at the wrecksite which is what we
have  been discussing above. 

High tensile steel would be expected to retain permanent magnetism for
longer than  mild steel or iron. I can find no reliable quantitative
data on this for steel plates used in ship building. One way to explore
further is too look at known wrecksites and see how the surface 
magnetic intensity can be understood.

Another complication is that for an old wreck, the direction of the earth's 
field will have changed during its period on the seabed. In the Liverpool 
area, magnetic north was  -21$^0$ from true north in 1877 and is now (2011) 
much closer (-2$^0$)  to true north.

This discussion of semi-permanent magnetism in ships shows that even more
uncertainty is  present in trying to predict the magnetic presence of a
shipwreck.  The semi-permanent magnetism can be represented by a dipole
distribution, just like the  induced magnetism. If one instead takes the
approach that the wreck is a magnetic dipole source of unknown strength, 
direction and distribution, then it is still possible to analyse this 
from data on the surface anomaly.

\section{Fitting surface data}
 \label{chap:fit}

Given an accurate set of data (position of measurement and value of 
magnetic intensity shift), it should be possible to deduce something
about the  distribution of magnetic material on the seabed. The general
case is discussed  in Appendix~C. If the surface data are sufficiently 
precise and extensive, it will be possible to interpolate them 
accurately and hence make a 2-dimensional Fourier transform of them.
This  allows in general, as discussed in Appendix~C, to evaluate
quantities of  interest which may help to deduce the nature of the
wreck. One example  considered in Appendix~C is the surface anomalous
magnetic energy distribution. This  is less sensitive (than the original
data on the magnetic field anomaly) to  the orientation of the dipoles
induced in the shipwreck. The surface distribution  of this quantity
thus tracks the underlying distribution of material more closely -
allowing the  orientation of the shipwreck to be inferred, for example.

As discussed above, if the only magnetic contribution is an induced 
magnetic dipole distribution (or a semi-permanent magnetism) in a known
direction, then  one can extract the distribution at the seabed directly
if one assumes that  all material is at the same depth. See Appendix~C1.

To go beyond this case, one needs to model the distribution (and
direction)  of the magnetic source and then fit the parameters of the
model to the  data taken at the surface. By fitting, I mean  minimising
the $\chi^2$ which is the sum of the square of
$(\mathrm{data-model})/\mathrm{error}$. This will require models of  the
shipwreck (unless detailed specifications exist) and I present  some
examples here that will be used in the comparison section.
See refs.~\cite{yangtze,scheldt} for other attempts at this.

\begin{figure}[t]
  \centering
  \includegraphics[width=.9\linewidth]{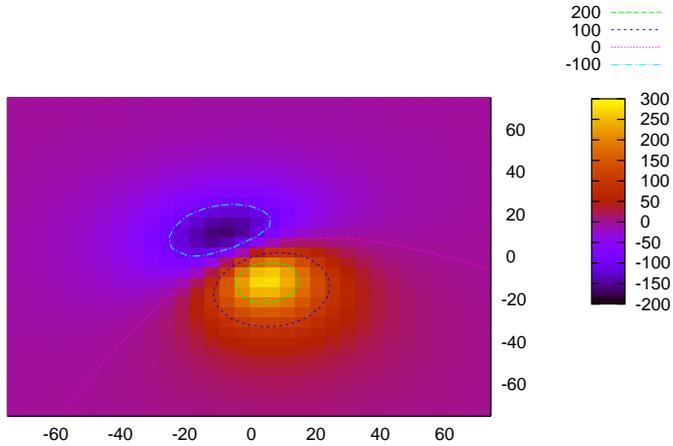}
  \caption{Magnetic field intensity anomaly (in nT) at the surface with
distances in metres from a 1 dipole fit to the observed data for the
wreckage of a  {\it Tower Base} in 25m in Liverpool Bay. 
  }
  \label{fig:trb3fit}  
\end{figure}

The simplest model is of a single magnetic dipole at the seabed.
Adjusting the position and orientation (see later discussion of {\it
Tower Base}) gives the surface  distribution shown in fig.~\ref{fig:trb3fit}

\begin{figure}[t]
  \centering
  \includegraphics[width=.9\linewidth]{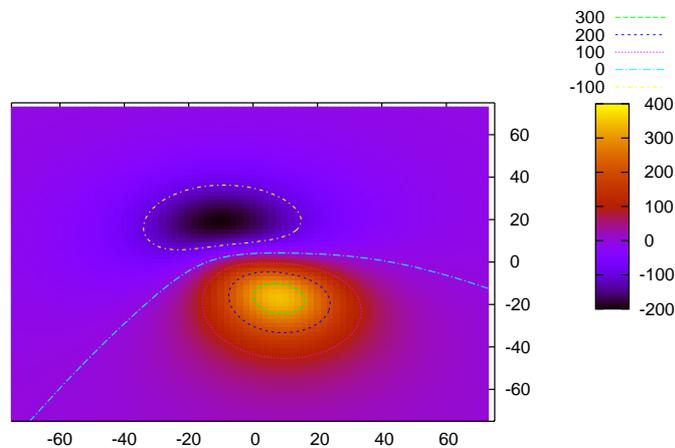}
  \caption{Magnetic field intensity anomaly (in nT) at the surface with
distances in metres for a model of the wreck of the {\it Ystroom} at 26m in
Liverpool Bay. Details are in the text.
  }
  \label{fig:yfit}  
\end{figure}

\begin{figure}[h]
  \centering
  \includegraphics[width=.9\linewidth]{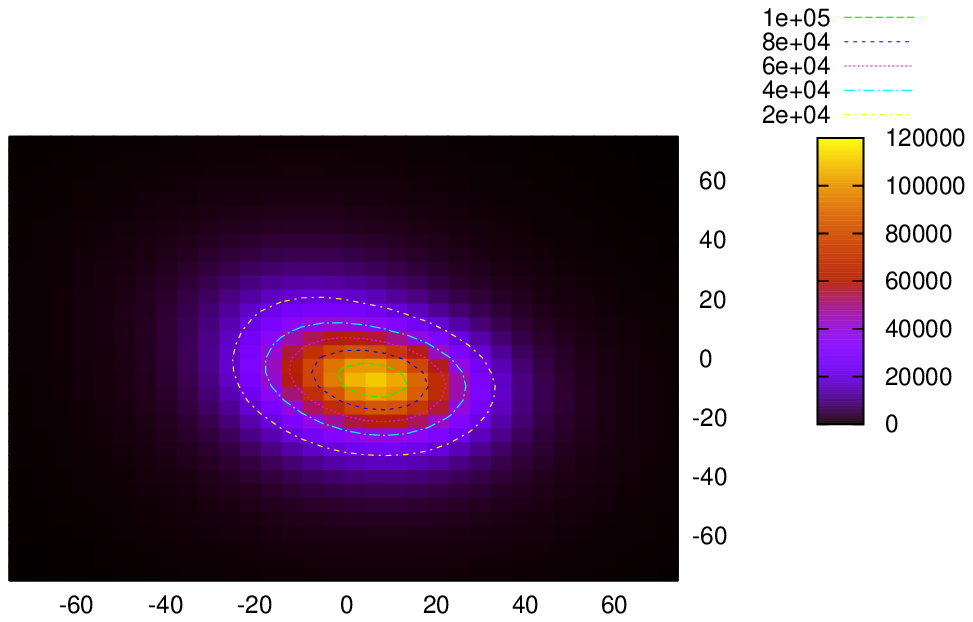}
  \caption{Magnetic energy density anomaly (in nT$^2$) at the surface with
distances in metres for a model of the wreck of the {\it Ystroom} at 26m in
Liverpool Bay. Details are in the text.
  }
  \label{fig:tubehh}  
\end{figure}

\begin{figure}[h]
  \centering
  \includegraphics[width=.9\linewidth]{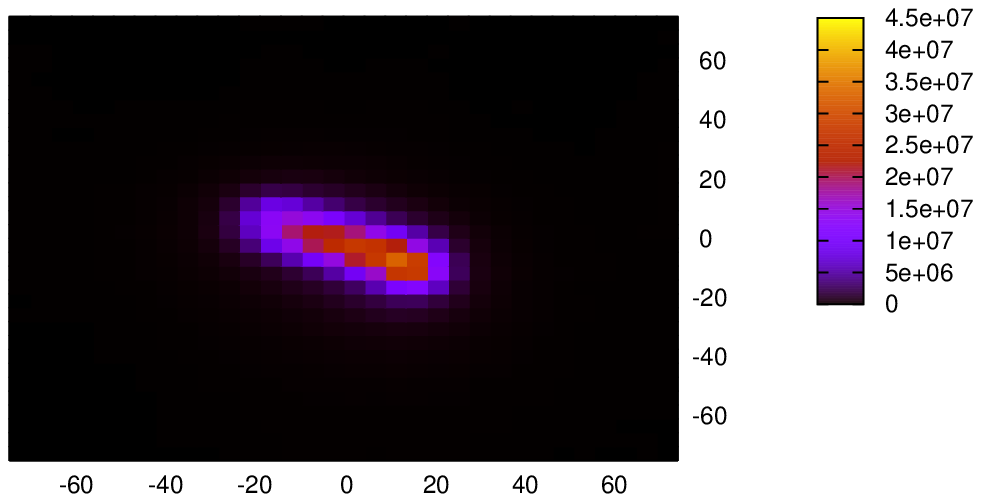}
  \caption{Magnetic energy density anomaly (in nT$^2$) at at height  of
5 metres above the seabed with distances in metres for a model of the
wreck of the {\it Ystroom} at 26m in Liverpool Bay. Details are in the
text.
  }
  \label{fig:tubehh5m}  
\end{figure}

 A model which reproduces many of the features of the data (see section
on wreck of {\it Ystroom}) is shown in  fig.~\ref{fig:yfit}. This model
has an open ended tube with rectangular  cross-section (8m wide, 3m
vertical) of length 40m and with plates having a combination of
thickness $t$ and relative permeability $\mu_r$ such that  $t
(\mu_r-1)=10$m. The tube lies 135$^0$-315$^0$.
  The dipoles induced in this model are evaluated following  the methods
described in Appendix~B4, yielding the result in fig.~\ref{fig:yfit}. 
 It is also of interest to evaluate the surface energy distribution for
this model since this quantity has a simpler interpretation. The result
is  shown in fig.~\ref{fig:tubehh}. This shows a distribution with a
maximum  which appears  less long than the 40m of the model, because of
saturation effects as  discussed in Appendix~B3. One can, in principle 
as discused in Appendix~C, also deduce the magnetic energy distribution on 
a level which is below the surface. In fig.~\ref{fig:tubehh5m} this 
distribution is shown for a level 5 metres above the seabed for the same 
model. This illustrates that the shape of the model is now more accurately 
reproduced.

\section{Examples of shipwrecks}

Here I illustrate these ideas with some practical cases from  Liverpool
Bay. More details on the wrecks involved are given in my
books~\cite{wlb}. Here the data  has been obtained  using my boat and
towing a proton magnetometer  behind. The boat's position is determined
by GPS. Typical boat speed through the water is  3.5 to 5 knots and the
measurements are taken after a 2 secs polarisation time. Corrections are
made for the length of the towing cable  and  for the current. Depth to
the seabed is noted by echo-sounder.

\subsection{Ystroom}

\begin{figure}[h]
  \centering
  \includegraphics[width=.9\linewidth]{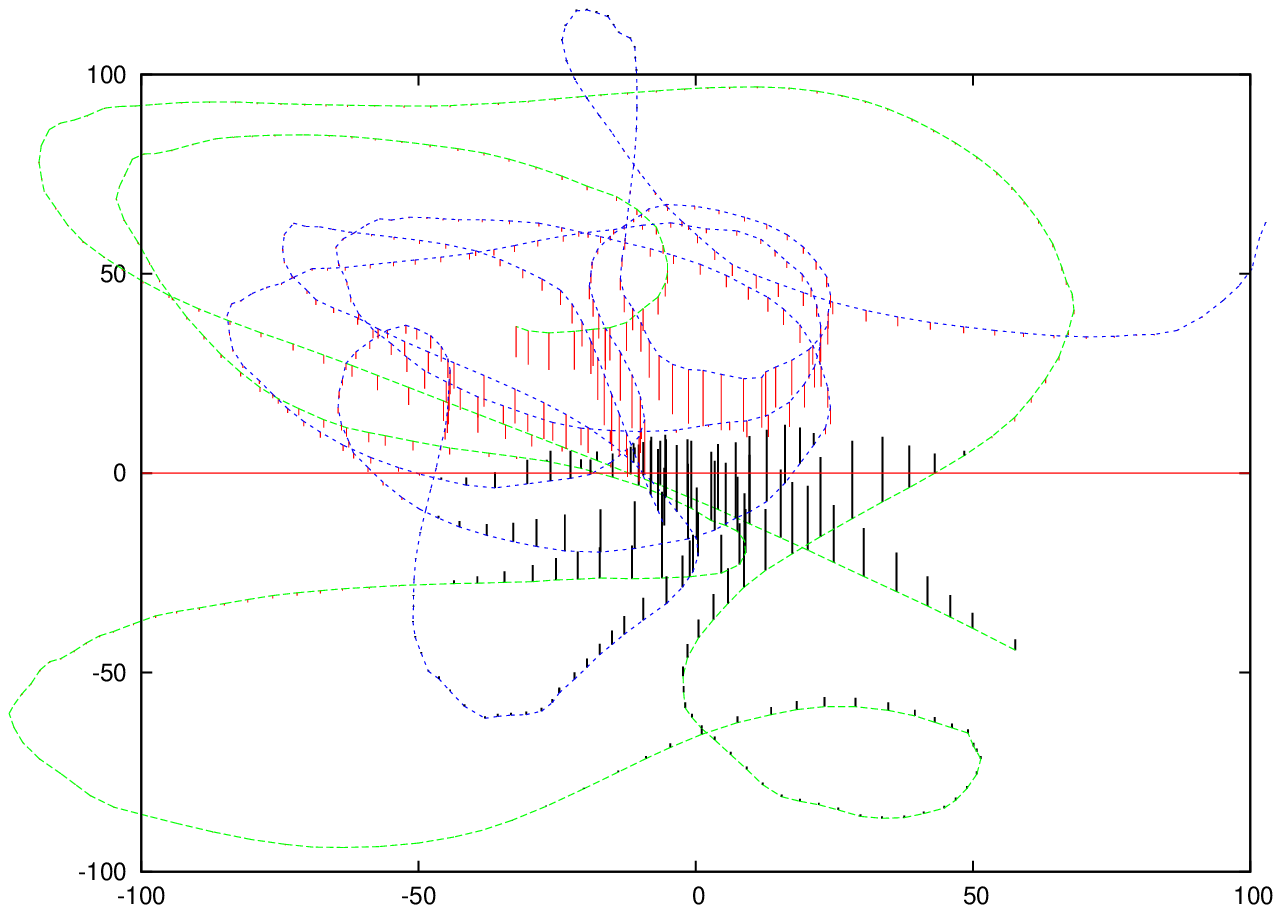}
  \caption{Raw magnetic field intensity anomaly data (in nT) at the
surface with distances in metres for the wreck of the {\it Ystroom} at
26m in Liverpool Bay. The dotted lines show the boat track (2 separate
tracks) while the vertical  bars (black for positive, red for negative)
show the magnetic anomaly. The data  have been corrected for the
difference between the measured (boat) GPS  position and the position of
the sensor towed behind.
  }
  \label{fig:ystroomraw}  
\end{figure}

\begin{figure}[h]
  \centering
  \includegraphics[width=.9\linewidth]{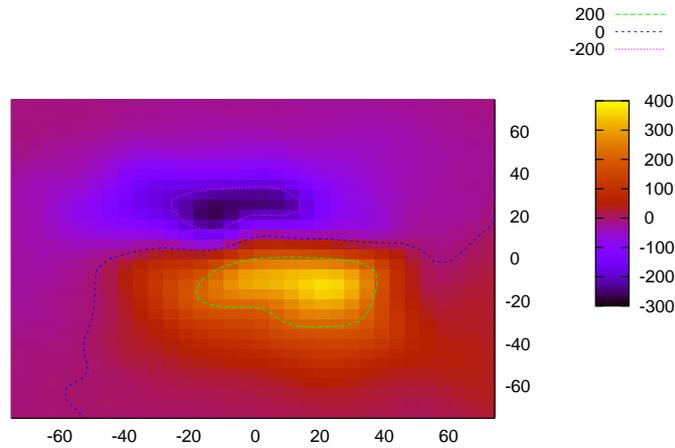}
  \caption{Magnetic field intensity anomaly (in nT) at the surface with
distances in metres for the wreck of the {\it Ystroom} at 26m in Liverpool Bay.
Details of smoothing used are in the text (box size of $300 \times 300$ metres 
with step size of 4.7 metres).
  }
  \label{fig:ystroom}  
\end{figure}

As a first example, I use data taken by me at the surface over the
wreck of the {\it Ystroom}.  This was a Dutch coaster of 400 gross tons
that sank from mine damage  in 1940. The wrecksite now has no wreckage
rising  more that a metre or so above the seabed - so the wreck has
collapsed and/or  become sanded in. For more detail about this wreck,
see the book `Wrecks of Liverpool Bay'~\cite{wlb}. The seabed was at 24
metres depth when I made the magnetic measurements and the wreckage is
mostly  under the seabed - so I use 26 metres as the average depth of
the wreckage. The raw data are shown in fig.~\ref{fig:ystroomraw} after 
correction for the cable length. They have been gridded using  both an
inverse power weighting and  planar interpolation  of a Delaunay
triangulation (see Appendix~D). Furthermore they have been smoothed  by
requiring high wave numbers to decrease at the expected rate as
discussed  in Appendix~C.  Similar results are obtained from  these
different  smoothing methods.  Typical results  are illustrated in
fig.~\ref{fig:ystroom}.

It is interesting to compare this measured surface anomaly directly with
that  obtained from the tube model and presented in
fig.~\ref{fig:yfit}. The overall distribution is rather similar.

\begin{figure}[h]
  \centering
  \includegraphics[width=.9\linewidth]{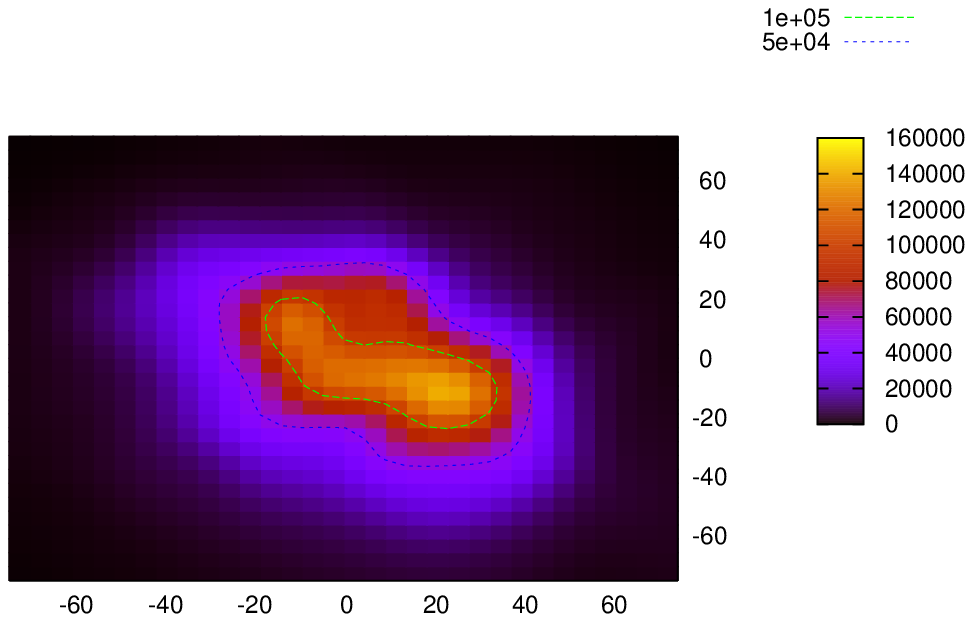}
  \caption{Squared magnetic field intensity anomaly (in nT$^2$) at the
surface with distances in metres for  the wreck of the {\it
Ystroom} at 26m in Liverpool Bay. Obtained from a 2-dimensional Fourier 
transform of the (interpolated) measured data using a $64 \times 64$ grid.
  }
  \label{fig:yhh}  
\end{figure}

Another way to process the data is to extract a quantity which is  more
simply related to the underlying material distribution.  From the
gridding discussed above, the two-dimensional Fourier transform was
evaluated  which allows to calculate the surface energy density - shown
in fig.~\ref{fig:yhh}. This directly shows the orientation of the wreck,
in  agreement with the comparison with the model distribution (shown in
figs.~\ref{fig:yfit} and \ref{fig:tubehh}).
  
\subsection{Tower Base}

\begin{figure}[t]
  \centering
  \includegraphics[width=.9\linewidth]{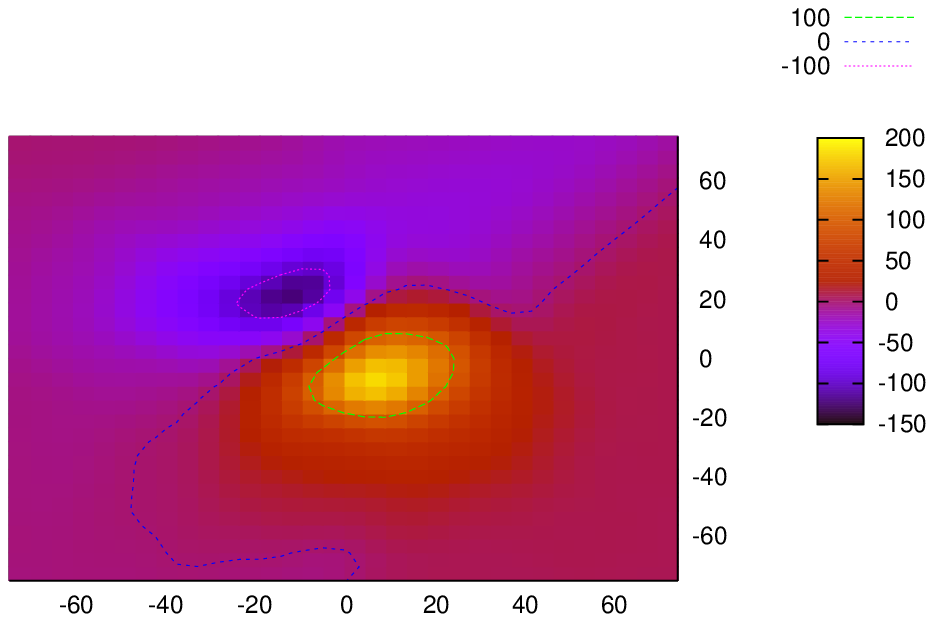}
  \caption{Magnetic field intensity anomaly (in nT) at the surface with
distances in metres  as observed (and smoothed) for the wreckage of a 
{\it Tower Base} in 25m in Liverpool Bay. 
  }
  \label{fig:trb3}  
\end{figure}

Another example studied is a {\it Tower Base} ~\cite{wlb} which is  the
dumped remains of the base of a world war II anti-aircraft fort.  This
is quite localised (about 35m long and 5m wide).  The (smoothed) surface
data are illustrated in fig.~\ref{fig:trb3} and a fit using one dipole
(so 3 dipole components plus the latitude  and longitude as parameters,
assuming the depth is 25 metres) has already been shown  in
fig.~\ref{fig:trb3fit}. This dipole has a horizontal component  pointing
to $326^0$ and a vertical downward component.

\subsection{Bouboulina}

 {\it Bouboulina} sank in 1867 after a boiler explosion while at anchor
in the Mersey. She was being fitted out as a Greek naval vessel after
being built at Liverpool as an  American Civil War blockade runner, {\it
Colonel Lamb}. For more detail see the book Lelia~\cite{lelia} about 
Liverpool's part in blockade running in the American Civil War.  She was
an iron and steel paddle steamer of 699 tons (gross). She split in two
after the boiler explosion and the stern section was salvaged.  The bow
section was reduced in height to cause less obstruction. Though 
charted, no sign of wreckage above the seabed is now visible (with
echo-sounder  or side-scan sounder). A substantial magnetic signal is
present, however. Since she was at anchor and the  current runs
$150^0/330^0$ at this location, her remains  would be expected to  lie
down current. At the time (near HW) I investigated this site, the  depth
to the seabed was 26 metres.

\begin{figure}[t]
  \centering
  \includegraphics[width=.9\linewidth]{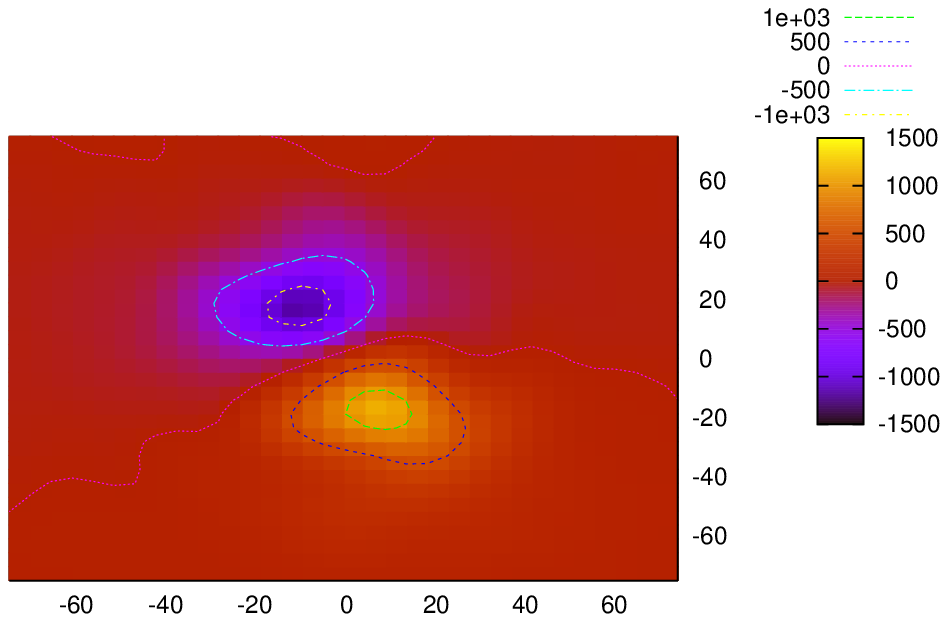}
  \caption{Magnetic field intensity anomaly (in nT) at the surface with
distances in metres  as observed (and smoothed) for the wreckage of 
{\it Bouboulina} (ex-{\it Colonel Lamb}) in 26m in the Mersey.
  }
  \label{fig:boub}  
\end{figure}

\begin{figure}[t]
  \centering
  \includegraphics[width=.9\linewidth]{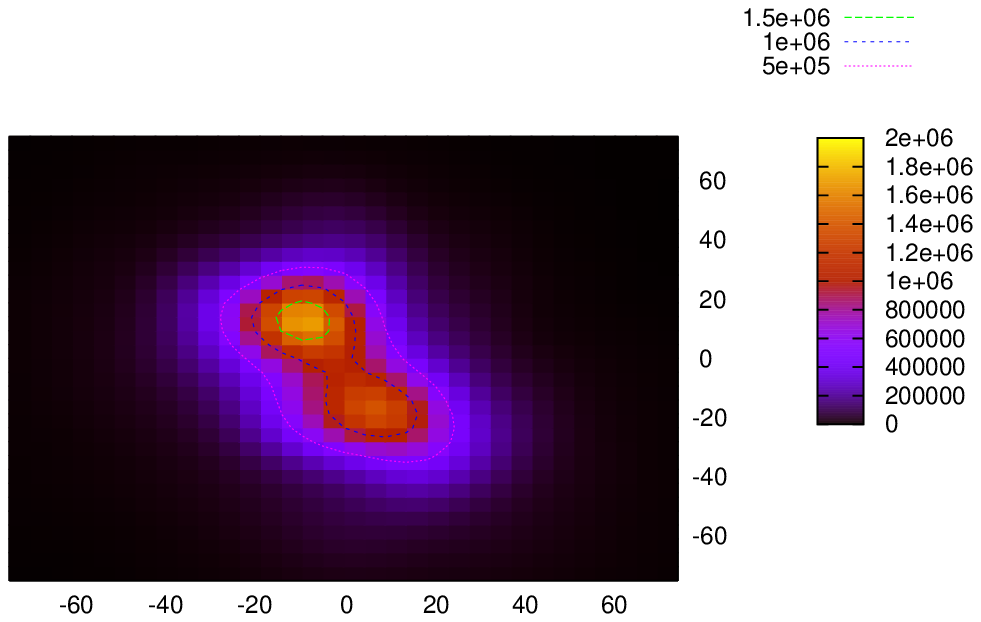}
  \caption{Magnetic energy distribution (in nT$^2$) at the surface with
distances in metres evaluated  for the wreckage of  {\it Bouboulina}  in
26m in the Mersey.
  }
  \label{fig:boubhh}  
\end{figure}

\begin{figure}[t]
  \centering
  \includegraphics[width=.9\linewidth]{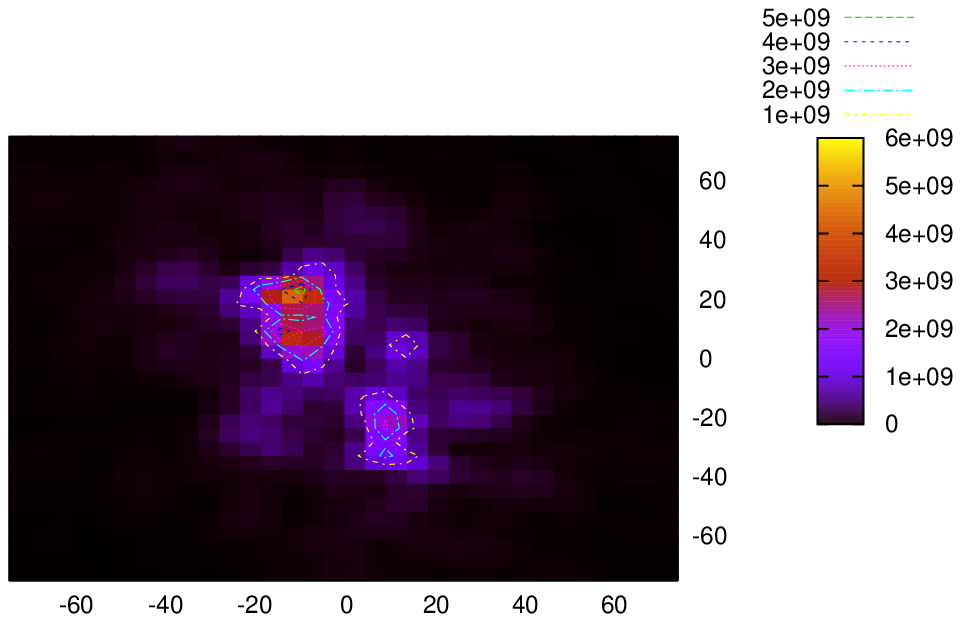}
  \caption{Magnetic energy distribution (in nT$^2$) at the surface with
distances in metres evaluated  at a level 5 metres above the seabed for
the wreckage of  {\it Bouboulina}  in 26m in the Mersey.
  }
  \label{fig:boubhh5m}  
\end{figure}

 The (smoothed) surface anomaly data are shown in fig.~\ref{fig:boub}.
After  taking the two-dimensional Fourier transform, the energy density
distribution  can be reconstructed - see fig.~\ref{fig:boubhh}. This
shows a distribution  running NW-SE as expected. The magnitude of the
anomaly signal (over 1000 nT)  is consistent with a large iron wreck and
the orientation fits with  expectation. As an example, I continue the
energy density distribution to a  depth corresponding to 5 metres above
the seabed. This sharpens the distribution  as shown in
fig.\ref{fig:boubhh5m}. This suggests that two areas of strong magnetic 
signal are dominant.  A fit using 2 dipoles (at fixed depth so 10
parameters)  gives a good description of the raw data - and confirms the
orientation as  NW-SE with a separation of about 30 metres.

The position I find for the wreckage is close (0.05nm) to that charted.
This is  certainly the remains of an American Civil War blockade runner.

\section{Conclusions}

A scattered distribution of roughly spherical ferrous material - such as
iron cannon balls - is likely to be magnetised along the earth's
magnetic field  and so can be modelled with that assumption. Methods
then exist which allow to  extract the seabed distribution from the
surface observations (see Appendix~C1  and C2).

Shipwrecks of iron or steel vessels are likely to have substantial iron
or  steel plates and these will be magnetised dominantly along the
surface of the  plate. Without a detailed knowledge of the construction
of the wreck (as  it lies on the seabed), it will be difficult to model
this accurately. As a guide,  I present some simple models: a
rectangular tube, a shoebox, etc.  These can  help to give an
understanding of the expected surface distribution. The rectangular tube
is shown to  give a surface distribution which is quite similar to that
of the wreck of  the {\it Ystroom}. In some cases, it may be  possible
to fit some of the parameters of such a model to the measured  surface
data. For example the {\it Tower Base}  can be  fitted adequately with
one dipole on the seabed while the {\it Bouboulina} can be well 
described by 2 dipoles. Note that these dipoles are found not to be aligned 
with the earth's field.

If no model assumption is justified, it is possible to use data
processing  (basically a two-dimensional Fourier transform) of the
measured surface  distribution to help visualise the seabed
distribution. I have presented  the case of the surface magnetic energy
which can be evaluated and which gives  a less biased picture of the
seabed magnetic source than the measured  anomaly itself (see
Appendix~C). Examples are given  for the wrecks of the coaster {\it
Ystroom} and the paddle steamer {\it Bouboulina}. These plots  give a
clear indication of the orientation and length of the wrecks. In
principle, it is possible to obtain this information at  a lesser height
above the wreck (than the surface) which will give even clearer
information, although this procedure can be  unstable unless the surface
data are very precise.


\section*{Appendix A: Surface intensity distribution}


 The discussion of static magnetic problems is well known~\cite{jackson}
but still  quite subtle:  the mathematics  assumes a knowledge of vector
calculus. There is also a review~\cite{blakely} focussing on geophysical
applications with some discussion of programming. A set of operations to 
analyse magnetic data are also available~\cite{magpick} and the manual 
(.pdf) can be found on the web.

 The magnetic flux field $B$ is measured in Tesla, while the  magnetic
field intensity $H$ is measured in amp/metre$^2$. Both of these fields
are  vectorial: they have a direction as well as an intensity. In free
space  $B=\mu_0 H$ with $\mu_0=4 \pi 10^{-7}$. This curious value of
$\mu_0$ arises since  the units used include the usual  electric units
(volts, amps) combined with  the metre-kilogram-second system.

In material, the relationship is $B= \mu H$  where  the relative 
permeability is  $\mu_r=\mu/\mu_0$. For most materials $\mu_r$ is close
to 1.0,  but ferromagnetic materials have large values (50-10000). The
situation  in a material can be considered as arising from an induced
magnetisation  (density of magnetic dipoles) $M$ where $H=B/\mu_0 -M$.
Thus the induced magnetism $M=(\mu_r-1)H$ which is a key relation for
this discussion - see Appendix~B. For strong  fields, the relationship
between $B$ and $H$ is complicated for ferromagnetic  materials (because
of hysteresis effects) but the earth's field is sufficiently   weak  
that a simple linear relationship is appropriate..

From Maxwell's equations, since there are no permanent electric
currents,  ${\bf \nabla \times H}=0$ and this allows to introduce a
magnetic scalar potential  $\phi$ with $ {\bf H=-\nabla} \phi$. This 
scalar potential satisfies the Laplace equation $\nabla^2 \phi=0$. 
Another Maxwell equation gives  ${\bf \nabla . B}=0$. These relations
imply the boundary conditions:  that tangential components of $H$ and
the normal component of $B$ are  continuous at a boundary.

 The field due to a collection of magnetic dipoles can most easily 
be evaluated using the magnetic scalar potential $\phi$.

 The magnetic scalar potential at ${\bf r}$ due to a (hypothetical) unit
magnetic source at ${\bf s}$ would be
  $$ V({\bf r})= 1/(4 \pi d({\bf r},{\bf s})) $$ 
 where $d({\bf r},{\bf s})$ means the distance from point ${\bf s}$  to
${\bf r}$. 
 For a magnetic dipole of strength and direction ${\bf p(s)}$ at ${\bf
s}$, the potential is 
 $$  \phi({\bf r})= {\bf p(s)}.\nabla_s \  1/(4 \pi d({\bf r},{\bf s})) $$ 
 The general case can be represented by a sum (over source locations
${\bf s}$) of such dipoles: 
  $$  \phi({\bf r})=\sum_s {\bf p(s)}.\nabla_s \ 1/(4 \pi d({\bf r},{\bf s})) $$ 
 The magnetic field can then be obtained from this potential
  $$  {\bf H(r)}= -\nabla_r \sum_s {\bf p(s)}.\nabla_s \ 
   1/(4 \pi d({\bf r},{\bf s})) $$ 
 
 This magnetic field must be added to the earth's magnetic field ${\cal
H}$ which is in the direction  given by direction cosines ${\bf
h}=(h_x,h_y, h_z)$.  Taking magnetic north as the
positive $y$-axis,  and the positive $z$-direction as upwards,  gives
$h_x=0$, $h_y=\cos(\psi)$ and $h_z=-\sin(\psi)$ where
 $\psi$ is the angle of dip, currently $67^0$ in the UK.
 The magnetic field intensity fluctuation observed at the surface will 
be dominated by the component $ {\bf h.H(r)}$ in practice, namely  
the intensity anomaly will be
 $\delta H(x,y)= -{\bf h}.\nabla_r \sum_s {\bf p(s)}.\nabla_s \  
1/(4 \pi d({\bf r},{\bf s}))$.

 This can be evaluated as 
 \be 
\delta H(x,y)=\sum_s (3{\bf h.R}\ {\bf p(s).R} -
        {\bf h.p(s)}\ R^2)/(4 \pi R^5)
 \label{eq:surf}
 \ee
 with ${\bf R=r-s}$.

 For a single magnetic source, one can choose  a coordinate system centred
 on that source. Then ${\bf s}={\bf 0}$ and ${\bf R}=(x,y,z)$ where $z$
will be the depth  of water if the source is at the seabed and  the
intensity  at the surface is evaluated. Furthermore, treating the
magnetic deviation by defining $y$ as magnetic north with  $x$ as
perpendicular (so approximately as longitude):
 $$ \delta H(x,y)=  \frac {3 (h_y y+ h_z z)(p_x x +p_y y +p_z z)  -
(h_y p_y +h_z p_z)(x^2+y^2+z^2)}{ 4 \pi(x^2+y^2+z^2)^{5/2}
} $$

A greater simplification occurs again if one assumes that the induced
magnetic dipole  ${\bf p}$ is directed also along the earth's magnetic
field direction.  Then ${\bf p(0)}={\bf h} p(0)$ where $p(0)$
is the dipole strength.
 So in this case
 $$ \delta H(x,y)=  p(0)\ \frac{3 (h_y y+ h_z z)^2
- (x^2+y^2+z^2)}{ 4 \pi (x^2+y^2+z^2)^{5/2} } $$

This expression is illustrated in fig.~\ref{fig:sample}. Note that 
since $B=\mu_r \mu_0 H$ with $\mu_r \approx 1.0$ and constant in air,
one can  quote the anomalous magnetic field as the ratio $\delta
H/\cal{H}$ or as the  same number in terms of the magnetic flux ratio
$\delta B /\cal{B}$. Here  $\cal{B}$ is the earth's magnetic flux
(48800 nT in the UK).

\section*{Appendix B: Induced dipole distributions}

\subsection*{Appendix B1: Spherical distributions}

A case that can be solved exactly: a sphere of ferromagnetic material of
radius $R$ with  relative permeability $\mu_r$ in a external magnetic
field $H$, assuming that $B=\mu H$ both externally and internally with
a  ratio of permeabilities $\mu_r$. The magnetic scalar potential 
satisfies Laplace's equation with appropriate boundary conditions on the
surface. Choose spherical polar coordinates with the external field
along the  polar direction, then the magnetic scalar potential inside
the sphere is
 $$ \phi=-hr \cos{\theta}$$
 and outside it is 
 $$ \phi=-H r \cos{\theta} + \frac{p}{4 \pi r^2} \cos{\theta} $$
 in this case, the induced dipole moment $p$ is located exactly at the
centre of  the material, this is not a general feature.

 Matching the tangential magnetic field ($H_t=-\frac{ \partial \phi} 
{r \partial \theta}$) and 
the radial magnetic flux ($B=\mu H_r = -\mu \frac { \partial \phi}
 {\partial r}$) at $r=R$ gives internal magnetic field
 $$ h = \frac{2}{\mu_r+1} H$$ 
 and a dipole strength of
 \be
 p= \frac{4 \pi R^3}{3}  (\mu_r-1) H \frac{3}{\mu_r+2}
 \label{eq:spher}
 \ee

 For comparison the induced dipole strength from the external field
alone is  $p=  \frac{4 \pi R^3}{3}  (\mu_r-1) H $ which  gives the same
as the exact result (eq.~\ref{eq:spher}) above for small $\mu_r-1$, as to
be  expected.

 Note also that as $\mu_r$ becomes big, the external magnetic field
becomes  normal to the sphere at its surface, since   the  tangential
component of $H$ which is
 $$-\frac{\partial \phi} {r \partial \theta}= H \sin{\theta}- p
\frac{\sin{\theta}} {4 \pi R^3}$$
  at the surface is zero when $p=H 4 \pi R^3$. This induced dipole
expression agrees with the exact expression  (eq.~\ref{eq:spher}) for 
large $\mu_r$. Moreover this induced dipole is basically independent of 
$\mu_r$ provided $\mu_r > 10 $.

There is a more general way to see why there is a maximum dipole moment 
induced. A dipole of strength $p$ aligned along the external field $H$
will produce  a field of strength $-p/(4 \pi r^3)$ transversly to its
location  and distance $r$ from it. This field is antiparallel to the
external field and cannot be so strong  as to reverse it,  so we
have $p < 4 \pi r^3 H$ where $r$ is the least distance from the dipole 
induced in the material to the exterior, transversly: namely $R$ in the
example above.
 This discussion makes it clear that the same reasoning can be applied
to a hollow sphere (a spherical shell) with the same maximum induced
dipole expression, where $R$ is now the  outer radius and $r$ the inner
radius. This case can actually  also be solved exactly, giving induced
dipole
 \be
 p=\frac{(2 \mu_r+1)(\mu_r-1)}{(2 \mu_r+1)(\mu_r+2)-(r/R)^3
 (\mu_r-1)^2} 4 \pi (R^3 -r^3) H  
 \label{eq:sphershell}
 \ee
 For small  $\mu_r-1$, this expression again agrees with the naive
expectation of the  volume of material times the  permeability, namely $
p \to 4 \pi (\mu_r-1) (R^3-r^3)H/3$

 This expression (eq.~\ref{eq:sphershell}) also again has the property
that it saturates at very large $\mu_r$  at a value $ 4 \pi R^3 H$ as we
argued above. However, it can be rewritten in a form  more useful for a
thin shell, with  $(r/R)^3=1-3t/R $ where $t$ is the  thickness of the
shell (if thin).
 \be
 p=\frac{(\mu_r-1) (2 \mu_r+1)} { 3 \mu_r + 2 (\mu_r-1)^2 t/R} 4 \pi R^2 t H
 \ee 



The criterion is now that saturation occurs when  $t \mu_r >> R$. For
plates used in ship-building  with  thickness $t \approx 0.03$ metres;
$\mu_r \approx 50 \to 1000$;  and size of structure $R \approx 10$
metres; then  the criterion is not fully satisfied: so one  does not
necessarily  have full saturation.

The induced dipole, because of the spherical symmetry, appears to be
located  at a point, the centre of the sphere. Thus the surface magnetic
anomaly is that  given by a point dipole (see Appendix~A and
fig.~\ref{fig:sample}). One can estimate the magnitude of the surface
anomaly in a typical case:  taking a sphere of radius 1.7 metres which
has volume about 20 m$^3$ and mass  of about 160 tons at depth 30
metres.  For typical $\mu_r$ values, there is saturation and this gives an
induced dipole $p=4 \pi R^3 H$ directed antiparallel to $H$ which
produces a magnetic field  of approximately $2 p /(4 \pi z^3)$ at the
surface where $z$ is the depth. This  gives an anomaly $0.00035 H$ which
is very small (about 17nT given the UK value of $H$). Note that a
spherical  shell of the same outer radius but of thickness 0.05 metre
with $\mu_r=800$ gives 94\% of the same surface  anomaly with only 9\%
of the volume, and thus mass, of ferromagnetic material.


What is learnt from these cases, is that the  dipole induced by the 
external field of the earth is modified by the magnetic field produced
by that  induced dipole itself. This can give rise to a saturation, the
induced dipole  does not increase further after $\mu_r \approx 10$ in the
case of a solid  object. For a thin shell, the criterion is that
saturation occurs when  $t \mu_r >> R$ and this inequality is often not 
satisfied  for ship building plates.

This discussion shows that it is not the volume of ferromagnetic
material that  determines the induced dipole strength, but rather the
spatial extent. So a relatively thin shell can have a similar induced
dipole to a  solid sphere.

In the case of the sphere and spherical shell, the symmetry of the situation 
ensures that the induced dipole is aligned with the external field. This 
is not generally the case, as I illustrate with a discussion of thin plates -
see Appendices~B3 and B4.

\subsection*{Appendix B2: a cylindrical rod and tube}


When a ferromagnetic object has infinite extent in one direction, the
problem  becomes effectively two-dimensional and thus easier to solve
exactly. One application  of this is to a long hollow cylinder, which
will be a useful model to compare with  computer methods to be discussed
later.

Consider first an infinite solid rod of radius $R$ with  relative
permeability $\mu_r$ and external magnetic field $H$ perpendicular to
the  axis.  Choose cylindrical coordinates, then the potential inside
the rod is  (using known solutions of the two-dimensional Laplace
equation)
 $$ \phi=-hr \cos{\theta}$$
 and outside it is 
 $$ \phi=-H r \cos{\theta} + \frac{p}{2 \pi r} \cos{\theta} $$

 Matching the tangential magnetic field ($H_t=-\frac{ \partial \phi} 
{r \partial \theta}$) and 
the radial magnetic flux ($B=\mu H_r = -\mu \frac { \partial \phi}
 {\partial r}$) at $r=R$ gives internal magnetic field
 $$ h = \frac{2}{\mu_r+1} H$$ 
 and a dipole strength per unit length, directed antiparallel to the 
external field $H$ 
 \be
 p =2 \pi \frac{\mu_r-1}{\mu_r+1} R^2 H 
 \label{eq:cyl} \ee

 For comparison the induced dipole strength from the external field
alone is  $p=  \pi R^2 (\mu_r-1) H $ which gives the same as the exact
result above for small $\mu_r-1$, as to be  expected.

 Again as $\mu_r$ becomes big ($ > 10$), the induced dipole strength
saturates  and the external magnetic field becomes  normal to the
cylinder at the surface.
 As before, a general argument shows why there is a maximum dipole
moment  induced. A dipole of strength $p$ aligned along the external
field $H$ will produce  a field of strength $-p/(2 \pi r^2)$ transversly
to its location  and distance $r$ from it. This field is antiparallel to
the external field and cannot be such as to reverse it,  so  $p <
2 \pi r^2 H$ where $r$ is the least distance from the dipole  induced in
the material to the exterior, transversly: namely $R$ in the example
above.
 
This discussion makes it clear that the same reasoning can be applied
to a hollow  cylinder with the same maximum induced dipole expression,
where $R$ is now the  outer radius. For a cylindrical shell of 1 mm
thick,  radius 1 metre with $\mu_r=999$, one finds this maximum is already
nearly  attained (83\%). So a thin cylindrical shell can have a similar 
induced dipole moment to a solid cylinder of the same outer radius. This
makes it clear that the volume of ferromagnetic material  is less
important than the spatial extent in regard to the maximum dipole 
induced.

The exact result is also known in this case (inner radius $r$, outer $R$):
 \be       p=\frac{2 \pi (\mu_r^2-1) (R^2-r^2)  }{
         (\mu_r+1)^2-(r/R)^2 (\mu_r-1)^2 } \label{eq:cylshell} \ee
 and again the criterion for saturation is that $t \mu_r >> R$ where the 
thickness $t=R-r$. In the above numerical example $t \mu_r=1$ and $R=1$
(both in metres), so there is partial saturation.

\subsection*{Appendix B3: flat plates}

 The boundary  conditions at a surface between water and ferromagnetic
material  are that (i)  tangential components of the magnetic field $H$
are continuous and (ii) normal  components of the magnetic flux  $B=\mu
H$ are continuous.  Ferromagnetic material has a very high permeability
$\mu$ and this implies that  the induced magnetisation is parallel to
the surface.  For a thin sheet of magnetic material, such as the plate
of  a ship,  the earth's magnetic field will induce  magnetic  dipoles
which have an orientation that lies in the plate. 

 For example, since the earth's magnetic field lies in direction 
pointing to magnetic north but downward angled,  then a horizontal plate
will be magnetised  with induced dipoles pointing north, while a
vertical plate oriented W-E will  have induced dipoles pointing
vertically downwards.

 The general case is less simple since the induced dipoles will themselves 
produce a magnetic field which needs to be taken into account consistently.
I present a general method for achieving this computationally below in 
Appendix~B4.

I first discuss  the relatively simple case of a thin infinite flat
plate of width $l$ and  thickness $t$ with relative permeability
$\mu_r$. If the external magnetic  field $H$ is in the plane of the
plate and perpendicular to the infinite extent,  one can solve the
problem analytically though not  easily. The key to understanding the
situation is that the dipole density, $(\mu_r-1)H$ per unit volume, 
induced by $H$ will be oriented along the width direction and be
uniform. The field  on the plate at distance $x$ from one edge caused by
that induced distribution will be given by
 $$ \frac{1}{2 \pi} H t (\mu_r-1) (\frac{1}{x} + \frac{1}{l-x} ) $$ 
 If this field (which is directed against the external field) is too big
then the  induced dipole distribution will be reduced. The largest
effect will  be near the edges when $x < t (\mu_r-1) / (2 \pi)$. This
can be modelled as an  effective reduction in the magnetised length from
$l$ to $l- t (\mu_r-1) / \pi$. For example with $t (\mu_r-1) =40$
metres, the length reduction  would be approximately 13 metres which is
indeed what one gets from the exact  solution when $ l > 40 $ metres.
For example, I find, from computer evaluation (with $t (\mu_r-1) =40$
metres again, see Appendix~B4 for details) that  a 100 metre long plate
(which would naively have  an  induced dipole 10 times that of a 10
metre long plate) actually  has an induced dipole 40 times bigger (since
the self consistency effects reduce the dipole for the 10m plate 
substantially).

 This analysis shows that the largest dipole will be induced
 \begin{itemize}
  \item when the component of the earth's magnetic field tangential to a
plate is greatest. 
  \item and when the length of the plate in that direction is longest.
 \end{itemize}

The earth's magnetic field  component vertically is  2.36 times as great
as that horizontally (north) in UK latitudes. In most shipwrecks the
vertical length of plates is relatively short, compared to the
horizontal length of  plates. The extra length compensates for the 
lower magnetic field component, this implies that horizontal induced
dipoles  are likely to be at least as important,  if not more so,
than vertical ones.

Here I make an estimate of the  absolute signal from a shipwreck.
Consider a horizontal steel flat plate on the seabed  of extent 40
metres by 10 metres and thickness $t=0.05$ metres (this is thicker than
a  typical plate, so is supposed to account for several). It has  volume
20 m$^3$ and a mass of about 160 tons. This mass is similar to that of a
small coaster. The induced  dipole moment is $p=20 (\mu_r-1) H_N$ where
$H_N$ is the component of the  earth's magnetic field in the magnetic
north direction (so $H \cos(\psi)$).  The field  produced by such a
dipole at the surface (horizontally, pointing south) above the  dipole
is  $p/(4 \pi z^3)$ and the anomaly (here a reduction  in intensity)
will be the component of this along the earth's magnetic field, so $p
\cos(\psi)/(4 \pi z^3)$. This gives a ratio of the  anomaly to the total
earth's field intensity of  $ 20 (\mu_r-1) \cos^2(\psi) /(4 \pi z^3)$
which gives, setting depth $z=30$ metres  and $\mu_r=800$,  a ratio of
0.007. This  is similar to what I expect from experience of actual
shipwrecks, namely a  signal of around 300nT in a total of 49000nT,
giving an observed ratio of 0.006.

 A computer evaluation of this case (actually with $(\mu_r-1)t=40$
metres, see Appendix~B4) gives a peak (actually a trough)  anomaly of
-372nT if the plate lies  N/S and -136nT if the plate lies E/W. The
largest signal, here a reduction, is not actually  directly above the
centre of the wreck but displaced to the north. The smaller signal  for
a plate lying E/W comes because the self-consistency is more stringent
in that case (the end effects are more serious if the  length in the
appropriate direction is only 10 metres not 40 metres as discussed
above). See fig.~\ref{fig:flat} for an illustration of the surface
distribution for the N/S case.

For a vertical plate of the same size and thickness (lying with centre 
at depth 30 metres, so half under the seabed), the peak anomaly is now 
positive at 763nT if  the plate is N/S and 599nT if it is E/W. These
signals are larger since the  earth's field has a relatively larger
vertical component (if the angle of dip is 67$^0$). 


Even though the earth's magnetic field lies in a (magnetic) north - vertical 
plane, induced dipoles with a E/W orientation can occur in general.
 For instance a similar vertical plate lying NE/SW has a net induced dipole 
in direction $(0.48,0.48,-0.74)$ so the horizontal component is directed 
along the direction of the plate - as it must be.

\subsection*{Appendix B4: Self consistent treatment of flat plates}

To determine the magnetic anomaly produced by ferromagnetic material,
one also  has to take into account, consistently,  the additional
magnetic induction caused by the  field produced by the induced dipoles
themselves. This is computationally feasible, if one knows the exact
location  and permeability of all magnetic material. Here I follow a
method recommended for use  with ships~\cite{plates}.

 A thin plate of ferromagnetic material (thin means that the thickness
is small  compared to other dimensions involved) can be treated
relatively easily as a thin shell.

 Treating the ferromagnetic shell as a series of elements with surfaces
$S_i$  (here taken as flat for simplicity of presentation) labelled
$i=1, \dots n$, one wishes to determine the magnetic dipoles ${\bf p_i}$
induced  on each element by the external field ${\bf H}$. Here the
induced dipoles will be treated as  located at the centre of each
surface element. Since only the tangential  component ${\bf H_t}$ of the
external field   contributes, integrating over the  surface of the
element $i$ gives the induced dipole:
 \be
 {\bf p_i}= k \int_{S_i} dS_i \ \ ( {\bf H_t - \nabla_t} \phi_i) 
 \label{eq:induce}
 \ee
 where $k=(\mu/\mu_0 -1)t$ represents the thickness of the material $t$
and the relative permeability $\mu_r=\mu/\mu_0$. 
 The second term on the right hand side represents the surface component
of the magnetic field created  by the induced dipoles themselves. This
is the term that I discussed qualitatively  in the approximate 
discussion above in Appendix~B3. 

Now $\phi_i$ is the magnetic static potential on the surface element
$S_i$ from all the induced dipoles (here treated as pointlike). So

 \be
 \phi_i=\sum_{j=1}^n {\bf p_j .} \ \nabla G_{ij} 
 \label{eq:phi}
 \ee

Here $G_{ij}$ is the Greens function ($1/(4 \pi |{\bf r}_i - {\bf
r}_j|)$) from the source at the  centre of element $j$ to a coordinate
in element $i$. Using Gauss' theorem on the surface of element $i$, one
can re-express the second term in eq.~\ref{eq:induce}, since
 $$ \int_{S_i} dS_i \ {\bf n.  \nabla_t}  \phi_i = 
   \int_i {\bf dl_i.n} \ \phi_i $$
 where ${\bf dl_i}$ is a length element of the closed boundary curve
encircling $S_i$ and is oriented normal (and outwards) to the curve but
tangential to $S_i$.  For a rectangular surface element, the component
in the direction (${\bf n}$) of the length of a side  will be given by
the difference of the integrals along the two sides normal to it. It is
convenient to describe the dipole moment ${\bf p_i}$ of each element $i$
by its  components in the two orthogonal directions of the sides (for a
rectangular element). 

 In this formulation, the unknowns are the $2n$ induced moments $p_i$
(which are  tangential to the surface elements so have only two
components each).
 Then one can solve for these induced dipoles $p_i$ by inverting a $2n
\times 2n$  matrix and  hence determine the anomalous magnetic field
($-\nabla \phi$) anywhere from them.

 I have written computer programs to evaluate these dipoles. As a check
one  can compare with known cases (such as a cylindrical shell) and one
can vary  the number of elements used and check for convergence of the
result as the  number is increased. The codes run almost instantly with
hundreds of elements.

 The result from one example is shown for the surface  magnetic anomaly
in fig.~\ref{fig:yfit} and for the surface energy anomaly in 
fig.~\ref{fig:tubehh}. This example is a rectangular cross-section tube
(8m wide, 3m vertical, 40m long)  with open ends and lying horizontally
with orientation NW-SE. The thickness  $t$ satisfies $t(\mu_r-1)=10$m. 
This was evaluated using a model with many small rectangular elements as
 described above (96 elements are sufficient).



\section*{Appendix C: analysing the surface field}

Here I first consider the general case: assuming only that the magnetic 
dipole sources are located in a limited area near the seabed - so away
from the surface.  In general, there might also be some geophysical
source of magnetism in  the rocks under the seabed - this is called a
`magnetic anomaly' on the  sea charts. Here I assume that the magnetic
dipoles are in a localised region -  as would be the case for a
shipwreck.

The anomalous magnetic field $H$ at the surface then can be derived from
a  potential $\phi$ (where ${\bf H}=-\nabla \phi$)  which satisfies the
Laplace  equation ($\nabla.\nabla \phi =0$). The observed intensity
shift $\delta H$ is given by $ {\bf h.H} $ where ${\bf h}$ is the
(known) direction  of the earth's magnetic field.

From the property that ${\bf \nabla.H}=0$, it follows, from the Gauss theorem, 
that 
 $$ \int dx dy\ H_z =0$$
 Also, since $H$ can be derived from a potential, it follows that
 $$ \int dx \ H_x= \int dy\ H_y =0$$ 
 Thus the observed anomaly satisfies 
 $$ \int dx dy\ \delta H =0$$
 This is a useful cross-check on any data taken.

 At large distance ($r$) from the shipwreck, the anomalous magnetic
field  (given by eq.~\ref{eq:surf}) will decrease as $1/r^3$ or faster.
This can also be  used to cross-check data.

A very useful way to analyse  data is with the  two-dimensional Fourier
transform:

   $$ \delta H(u,v,z)= \int dx dy \ e^{-iux - ivy} \delta H(x,y,z) $$

Here $u$ and $v$ will be referred to as wave-numbers. The inverse of
this transform is given by:

   $$ \delta H(x,y,z)= \frac{1}{4\pi^2} \int du dv\ e^{iux + ivy} \delta H(u,v,z) $$

In terms of the potential and its Fourier transform $\phi(u,v,z)$, we
have  
   $$ \phi(x,y,z)= \frac{1}{4\pi^2} \int du dv e^{iux + ivy}  \phi(u,v,z) $$
and the Laplace equation constraint implies that
 \be  
 \frac{\partial^2 \phi(u,v,z)}{\partial z^2}=(u^2+v^2) \phi(u,v,z)
 \label{eq:laplace}
  \ee

 Defining $k^2=u^2+v^2$, then  $\partial \phi(u,v,z)/\partial z=-k
\phi(u,v,z)$ since  $\phi$ must decrease at large values of $z$. This
enables  evaluation of the gradient of $\phi$ and hence  the magnetic
field associated with it:

   $$ {\bf H}(u,v,z)= (-iu,\ -iv,\ k) \phi(u,v,z) $$

Thus the potential can be extracted from the observed magnetic field anomaly
which is measured at the surface ($z$ is the height of the surface above the 
seabed):

   $$ \phi(u,v,z)= \frac{\delta H(u,v,z)}{-iuh_x-ivh_y+kh_z}$$

Note that  when $u=0$ and $v=0$, then $k=0$ and hence the division is 
ill-defined. This case corresponds to the sum of $\delta H$ over all 
space which must be zero as discussed above. So there is a $0/0$
situation: the average value of $\phi(x,y,z)$ over $x$ and $y$  is not
determined.  This is resolved by the fact that an  overall constant
change to $\phi(x,y,z)$ is not physically important -  only differences
of potential values are significant. Moreover, in practice,  one may not
need to extract the potential itself, but quantities such as magnetic
field components derived from differences of it.

In order to illustrate the significance of the two-dimensional Fourier 
transform, I present the transform of the general case given in 
eq.~\ref{eq:surf} in Appendix~A. Since the transform of
$(x^2+y^2+z^2)^{-1/2}$ is  $ 2 \pi \exp(-kz) / k$ where $k^2=u^2+v^2$,
one can evaluate 
$$
 \delta H(u,v,z)= \sum_s \ {\bf h.K} \ {\bf p_s.K}\  e^{-ius_x-ivs_y} 
  e^{-k(z-s_z)} /( 2 k)
$$
 where the wave-number vector ${\bf K}=(-iu,-iv,k)$ and the sum is over 
dipoles of strength and direction ${\bf p_s}$ located at $(s_x,s_y,s_z)$

The most noticeable feature of this expression is the factor $\exp(-kz)$
which  can be deduced on general grounds from the Laplace equation as
shown in  eq.~\ref{eq:laplace}. This factor  controls the rate at which
the Fourier transform decreases for large  wave-number. Large
wave-number is  related to finer structure: so this can be seen  as the
quantitative expression of the fact that the surface distribution  will
be smoother, the deeper is the wreck.
This can be used to smooth data - by requiring that the wave-number 
distribution falls off in magnitude as $k \exp(-kz)$ at large $k$.

One can also estimate the range of important wave-numbers since $k
\exp(-kz)$ is  maximum at $k \approx 1/z$.  If a total spatial size $L
\times L$ with  grid step interval $s \times s$ is used to process the
data, the range of sizes of wave-number  will be from $2 \pi/L$ to $ 
\pi / s$. Ideally the minimum non-zero  wave number ($ 2 \pi /L$) should
be several times smaller than the  typically important wave-number
($1/z$). This implies $L > 2 \pi z$. So  for a depth $z$  of 25 metres
the box size used should be of linear  size greater than 160 metres. 

As well as the potential (and derived quantities) at the surface (height
 $z$ above the wreckage), one can evaluate the potential at a lesser
height  ($z_0 < z$ and $z_0$ greater than the vertical coordinate of any
part of the wreckage itself) above the wreck since 

$$ \phi(u,v,z_0)=e^{k(z-z_0)} \phi(u,v,z)$$

Since $k^2=u^2+v^2$, this will enhance contributions with larger  $u$
and $v$ which correspond to more detailed structures in $x$ and $y$.
Thus, at least in principle, one can continue the surface data down to a
small height $z_0$,  such that $z_0$ is above every part of the wreck,
which will show up the detailed spatial structure more clearly. This, 
of course, relies on having sufficient precision in the data taken at
the  surface. Only in that case will the high wave-number components 
reproduce the required rate of drop off  (as $\exp(-kz)$). As an
estimate, if the data are available with a step  size of $s$ metres,
then the enhancement $\exp{k(z-z_0)}$ will be   $ \exp{\pi (z-z_0)/s}$
for the largest wave-numbers. If $z-z_0 > s$, this large enhancement
factor  will multiply the very small contribution from such large
wave-numbers, but  the numerical procedure of obtaining the Fourier
transform for such large  wave-numbers may introduce errors that
de-stabilize the analysis.
 Hence possible instabilities can arise in continuing to lower depths by
 an amount greater than the   spatial resolution $s$.

Returning to the discussion of an appropriate combination to evaluate, 
once the potential has been evaluated,  one suitable combination is 
${\bf H.H}$, which is proportional to the total anomalous magnetic field
energy  at the surface. This quantity is less sensitive to the
orientation  of the magnetic dipoles at the seabed than the observed 
anomaly itself ($\delta H=\bf{ h.H}$). Indeed, from a dipole
${\bf p}$ at  relative position ${\bf R}$, 
 $$ {\bf H.H}= p^2 \frac{3 \cos^2 \theta + 1}{16 \pi^2 R^6}$$ 
  where $\theta$ is the angle between the directions of ${\bf p}$ and
${\bf R}$. Thus the signal is always positive and varies by a factor of,
at most,  4  for any $\theta$  at fixed distance $R$. This makes the
surface signal easier to interpret, compared to  $\delta H$ which can
have sign changes, etc. Furthermore, this quantity will show quite a
localised  surface distribution: for instance, a point source at depth
$z$ will  spread out on the surface so that the peak  lies above the
point source and the signal drops to 50\% within a  circle of radius
$z/2$ (for the case of a vertical dipole).

This can be illustrated with the rectangular tube model of a shipwreck 
introduced in Appendix~B4, yielding the result in fig.~\ref{fig:yfit} 
for the surface anomaly.  For the same model, the surface energy
distribution is  shown in fig.~\ref{fig:tubehh}. This is closer to the
underlying material distribution  but shows a distribution with a maximum 
which appears  less long than the 40m of the model, because of
saturation effects as  discussed in Appendix~B3. One can, in principle 
as discused above, also deduce the magnetic energy distribution on  a
level which is below the surface. In fig.~\ref{fig:tubehh5m} this 
distribution is shown for a level 5 metres above the seabed for the same
 model. This illustrates that the shape of the model is now more
accurately  reproduced.

As a practical example, I show the distribution of ${\bf H.H}$ on the
surface for the  wreck of the {\it Ystroom} in fig.~\ref{fig:yhh}. This
shows a distribution  lying NW-SE, in agreement with the result of
comparing models with the  observed surface anomaly distribution.

Another combination, $\sqrt{\bf H.H}$, has an even smaller dependence
on the  orientation of seabed dipoles (a factor of 2 at maximum) and
would be a  possible quantity to present. For a dipole at the seabed,
the combination ${\bf H.H}$ has a more peaked spatial distribution at the 
surface, and this is why I favour it.

Using the above formalism to evaluate ${\bf H.H}$ at a lesser height
($z_0$) above the  seabed, gives rather noisy and unstable results
(unless  very accurate data exist or unless the data are processed
taking  account of the expected decrease with wave number) if evaluated
more than a few metres below the surface, as expected from the
discussion  above.



 \subsection*{Appendix C1: deducing the bottom distribution}

Here I consider a simple model: all the ferro-magnetic material  lies at
the same depth below the surface ($z$) and it is all magnetized  along
the same direction ${\bf a}$.  Then ${\bf p(s)}={\bf a} p(s)$ where
$p(s)$ is the dipole strength.

 This assumption could be appropriate to  a wreck consisting of a
scattered collection of iron cannon balls, for example. They will  then
all be magnetised along the earth's magnetic field direction which is
the  special case  that the direction ${\bf a}$ is that of the
earth's magnetic field itself (${\bf h}$).

From Appendix~A the observed deviation in magnetic field intensity will be
given by eq.~\ref{eq:surf}
 $$ \delta H(x,y)=\sum_s (3{\bf h.R}\ {\bf p(s).R} -
        {\bf h.p(s)}\ R^2)/(4 \pi R^5) $$
 where one introduces a magnetic dipole density $\rho(s_x,s_y)$ to describe
the  seabed distribution and makes use of the fixed direction of the
induced dipoles. So
 $$\delta H({x,y})=\int ds_x ds_y \
  (3{\bf h.R}\ {\bf a.R} - {\bf h.a}\ R^2)\rho(s)/(4 \pi R^5)$$ 

This expression has the form of a convolution
 $$ \delta H(x,y) = \int ds_x ds_y \ G(x-s_x,y-s_y) \rho(s_x,s_y) $$
where  $G$ is a known  function since $z$ has been fixed at the depth
below the surface and ${\bf h}$ and ${\bf a}$  are both known.
 Our aim is to measure $\delta H(x,y)$ and thence deduce
$\rho(s_x,s_y)$. This  can be achieved by taking the 2-dimensional
Fourier transform of the expression, using

   $$ G(u,v)= \int dx dy \ e^{-iux - ivy} G(x,y) =
 {\bf h.K} \ {\bf a.K}\  e^{-kz} /( 2 k)
 $$

   $$ \delta H(u,v)= \int dx dy \ e^{-iux - ivy} \delta H(x,y) $$

Then  $\rho(u,v)=\delta H(u,v)/G(u,v) $ and so one can reconstruct

  $$ \rho(s_x,s_y)= \frac{1}{4\pi^2} \int du dv
           \ e^{ius_x + ivs_y} \rho(u,v) $$

However, as noted in the previous section, the correction  $1/G(u,v)$
will be  very large  for large wave number ($k$) because of the
exponential  factor $\exp(kz)$. This will render the approach unstable
unless  very precise data are available to give an accurate Fourier
transform of  $\delta H(x,y)$.

\subsection*{Appendix C2: Return to pole derivation}

At the magnetic north pole, where the magnetic field is vertical, the
magnetic  intensity pattern on the surface above a magnetic item can be
interpreted more easily.  An isolated magnetised region will show an 
enhancement directly above it and a decrease all around. So, roughly,
the area  of increased magnetic intensity will correspond to the
presence of iron below.  This makes interpretation of the surface anomaly 
easier.

However, Britain is not at the north magnetic pole. With some key
assumptions, it is possible  to process data so that they are in the
form they would take if they were measured at the  north pole. This is
known as `Return  to pole'.

 As discussed in Appendix~A, the  magnetic field intensity change
observed at the surface will  be given in terms of the magnetic scalar
potential $V$ from a (hypothetical) unit point magnetic source at
${\bf s}$ by
 $$\delta H(x,y)= -{\bf h}.\nabla_r \sum_s {\bf p(s)}.\nabla_s \  
V({\bf R})$$
 with ${\bf R=r-s}$ where ${\bf r}=(x,y,z)$ is the coordinate  of
observation and ${\bf s}$ is the location of the magnetic dipole source
(of strength and direction given by ${\bf p(s)}$.

 The key assumption is that the magnetic dipoles induced are all aligned
 along the earth's magnetic field direction given by ${\bf h}$.
Again (as in Appendix~C1) this assumption  is most appropriate for a
scattered collection of wreckage - rather than a ship with iron plates.
So ${\bf p(s)}= {\bf h} p(s)$ where $p(s)$ is the dipole strength
at ${\bf s}$ and  ${\bf h}$ is independent of ${\bf s}$. One may
include this  factor of $p(s)$ within the sum over sources defining a
scalar potential $\Phi$ (note not the same as $\phi$ introduced  in the
introduction to Appendix~C):  $\Phi=\sum_s p(s) V({\bf R})$

 Now consider the Fourier transform (in the two horizontal spatial coordinates 
$x$ and $y$ of ${\bf r}$). Here $z$ is the vertical coordinate of ${\bf r}$.
 $$ \Phi(x,y,z) = \frac{1}{4\pi^2} \int du dv \ e^{iux+ivy} \Phi(u,v,z)$$

 Since the  potential satisfies Laplace's equation $\nabla. \nabla \Phi=0$.
 This implies that 
 $$ (u^2 + v^2) \Phi(u,v) = \frac{\partial^2 \Phi(u,v)}{ \partial z^2}
= k^2 \Phi(u,v) $$

 This allows the gradients to be evaluated, giving the magnetic
field intensity anomaly 
 $$\delta H(x,y,z) = \frac{1}{4\pi^2} \sum_s \int du dv \ ((-iu,-iv,k).{\bf h})^2  p(s)
 e^{iux+ivy} \Phi(u,v,z)$$

 In contrast the field at the surface when the earth's magnetic field 
is vertical (at the pole) is given by ${\bf h}=(0,0,1)$. Hence
 $$ \delta H_{pole}(x,y,z) =\frac{1}{4\pi^2} \sum_s \int du dv \ k^2  p({s})
 e^{iux+ivy} \Phi(u,v,z)$$

Within our assumptions, these two expressions are related by a factor of 
$k^2/(-ih_x u-i h_y v + h_z k)^2$ where $k^2=u^2+v^2$.
This factor is independent of the source ${\bf s}$ and so one does not
need  to know the distribution of magnetic material in order to make the
correction.

Thus the $u,v$ Fourier transform of the measured intensity $\delta
H(x,y,z)$ can  be corrected by this factor where ${\bf h}$ is
known. For simplicity,  neglecting the magnetic deviation  and choosing
$y$ as latitude,  $h_x=0$, $h_y=\cos(\psi)$,
$h_z=-\sin(\psi)$ where $\psi$ is the angle of dip.  

\section*{Appendix D: Interpolating point data}

 Assume one has logged data (such as the magnetic intensity at the
surface)  at a set of locations. Data values are $v(x_i,y_i)$ with
$i=1,\dots n$ representing the  $n$ values logged at locations
$(x_i,y_i)$. One needs to interpolate these values for several purposes:
 to make a Fourier transform, to draw a smooth surface on a plot, etc.

 One straightforward way to do this is to make a weighted average of the data.

 $$ v(x,y) = \frac{1}{W} \sum_{i=1}^{n} v(x_i,y_i) w(x-x_i,y-y_i)$$ 
 with  $ W =  \sum_{i=1}^{n}  w(x-x_i,y-y_i)$.

A sensible weight to use is one that emphasises close values preferentially:
 $$w(x-x_i,y-y_i)= \frac{1} {((x-x_i)^2 + (y-y_i)^2 )^p}$$
 Choosing $p=1$ corresponds to using an inverse square distance as
weight. Choosing very large $p$ corresponds to using the nearest data
point. A reasonable compromise  occurs using an  inverse power such as
$p=2$. 

Note that this approach is appropriate for interpolating data but it 
will  not reproduce the property that a magnetic field intensity should
drop  off with a known power (the inverse cube)  at large distances
unless this condition is explicitly imposed.

 A more sophisticated way to interpolate data is to introduce a
triangular  network in $x,y$ and use planar interpolation (in $x,y,v$)
in each such triangle. The usual  choice of triangles is that
corresponding to the Delaunay triangulation (this  has no other vertex
points in the circumcircle of any triangle). To set up such a
triangulation, one must avoid repeated vertices and  sets of three or
more vertices lying on a line. It can also be helpful to add a few
vertices near the  boundary (with $v$ value zero) if there are not
measurements in that region.


\end{document}